\documentclass[11pt,a4paper,american]{article}
\pdfoutput=1
\usepackage{lmodern}

\usepackage[T1]{fontenc}
\usepackage[latin9]{inputenc}
\usepackage{mathrsfs}
\usepackage{amsmath}
\usepackage{amssymb}

\makeatletter

\pdfpageheight\paperheight
\pdfpagewidth\paperwidth

\usepackage{jcappub}
\numberwithin{equation}{section}

\pdfoutput=1
\renewcommand\[{\begin{equation}}
\renewcommand\]{\end{equation}}

\makeatother

\usepackage{babel}
\begin{document}

\title{Recovering $P(X)$ \\ from a canonical complex field}

\subheader{LPT-Orsay-18-89}

\author[a,b]{Eugeny Babichev,}

\author[c]{Sabir Ramazanov,}

\author[c]{and Alexander Vikman}

\affiliation[a]{\em Laboratoire de Physique Th\'eorique, CNRS, }

\affiliation{\em  Univ. Paris-Sud, Universit\'e Paris-Saclay, 91405 Orsay, France\\
 }

\affiliation[b]{\em UPMC-CNRS, UMR7095, Institut d'Astrophysique de Paris, ${\mathcal{G}}{\mathbb{R}}\varepsilon{\mathbb{C}}{\mathcal{O}}$}

\affiliation{\em 98bis boulevard Arago, F-75014 Paris, France\\
}

\affiliation[c]{\em CEICO-Central European Institute for Cosmology and Fundamental Physics, }

\affiliation{\em Institute of Physics of the Czech Academy of Sciences,\\
Na Slovance 1999/2, 18221 Prague 8, Czech Republic\\
}




\abstract{We study the correspondence between models of a self-interacting canonical complex scalar field and $P(X)$-theories/shift-symmetric k-essence.  Both describe the same background cosmological dynamics, provided that the amplitude of the complex scalar is frozen modulo the Hubble drag. We compare perturbations in these two theories on top of a fixed cosmological background. 
The dispersion relation for the complex scalar has two branches. In the small momentum limit, one of these branches coincides with the dispersion relation of the $P(X)$-theory. Hence, the low momentum phase velocity agrees with the sound speed in the corresponding $P(X)$-theory. 
The behavior of high frequency modes associated with the second branch of the dispersion relation depends on the value of the sound speed. In the subluminal case, the second branch has a mass gap. On the contrary, in the superluminal case, this branch is vulnerable to a tachyonic instability. We also discuss the special case of the $P(X)$-theories with an imaginary sound speed leading to the catastrophic gradient instability. The complex field models provide 
with a cutoff on the momenta involved in the instability.}
\maketitle

\section{Introduction and Summary}

Models with non-canonical kinetic terms of a scalar field are nowadays
quite common in modified gravity and cosmology. These theories are used to model the early and late
time acceleration of the Universe as well as Dark Matter, for a recent review see~\cite{Clifton:2011jh,Joyce:2014kja}. 
However, general derivative interactions instigate a number of pathologies: various types of instabilities and singularities.  
Many of these theories can avoid ghost and gradient instabilities for physically interesting solutions, but they remain vulnerable to caustic formation~\cite{Babichev:2016hys}\footnote{The appearance of caustics is more evident for dust-like theories, see Refs.~\cite{Frolov:2002rr,ArkaniHamed:2005gu,Blas:2009yd,Mukohyama:2009tp}.} (see also Refs.~\cite{Mukohyama:2016ipl, deRham:2016ged, Pasmatsiou:2017vcw,Tanahashi:2017kgn}). Presence of those singularities  appeals for a modification of these theories at short scales.  

In this work, we focus on the subclass of k-essence models~\cite{ArmendarizPicon:1999rj, ArmendarizPicon:2000dh, ArmendarizPicon:2000ah,Chiba:1999ka} 
with the shift symmetry, $\varphi\rightarrow\varphi+c$, so-called $P(X)$-theories. Here $X$ is the kinetic
term of the scalar field $\varphi$, namely $X=\frac{1}{2}(\partial\varphi)^{2}$. Notably $P(X)$-theories describe low energy dynamics of zero-temperature superfluids, Refs.~\cite{Greiter:1989qb,Son:2002zn,Son:2000ht,Alford:2012vn,Berezhiani:2015bqa}.
Like generic k-essence, $P(X)$-theories also develop caustic singularities~\cite{Babichev:2016hys}:
characteristics of the equations of motion intersect at some point, and
the second derivatives of the field $\varphi$ blow up. 

In Ref.~\cite{Babichev:2017lrx} the caustic free
completion of $P(X)$-theories has been proposed. The idea is to
promote the scalar field $\varphi$ to the phase of a canonical
complex scalar field $\Psi=|\Psi|e^{i\varphi}$ with a self-interacting potential. A link between k-essence and
the complex scalar field has been pointed out earlier in Refs.~\cite{Bekenstein:1988zy,Bilic:2008zk,Bilic:2008zz,Tolley:2009fg,Elder:2014fea}. It is straightforward to show that the phase of the field $\Psi$ indeed reproduces dynamics of the field $\varphi$ in a $P(X)$-theory, if one switches off dynamics of the amplitude $|\Psi|$, see Refs.~\cite{Bilic:2008zk, Bilic:2008zz} and Sec.~\ref{sec:From-fluid-to-U(1)}. 

Near (would be) caustics the behavior of the $P(X)$-theory and that of the canonical scalar field are clearly different: the former develops singularities, while the latter remains regular. 
Instead, in cosmology, far away from the regime, where caustics are formed, we prove that the background dynamics of $P(X)$-theory is recovered from the model of the canonical complex scalar field under certain conditions. Formulating these conditions is one of the goals of the present work. To fulfil these conditions one arranges a configuration of the complex scalar such that its amplitude is constant modulo the cosmological drag. 
In Sec.~\ref{sec:HomoCosmo}, we show that such a configuration exists, provided that the Hubble rate is slow 
relative to the phase time derivative $\dot{\varphi}$.

What is more important, in this work we recover the dispersion relation of the $P(X)$-theory from that of the canonical complex field. 
We accomplish this in Sec.~\ref{sec:Dispersion-relations-Cosmo} by studying the propagation of perturbations on a homogeneous and isotropic cosmological background in the test-field approximation, i.e., neglecting the perturbations of the metric.

The complex scalar field propagates two degrees of freedom (d.o.f.). Correspondingly there are two branches in the dispersion relation. In the infinite momentum limit, $k\rightarrow\infty$, both d.o.f. have a standard dispersion relation 
$\omega^{2}=k^2$. This is expected, as we deal with a canonical field. Note that restoring the speed of propagation equal to unity at small scales is sufficient for resolving caustic singularities of the $P(X)$-theory. In this work we demonstrate that in the limit of small
momenta, $k\rightarrow 0$, one branch of the dispersion relation recovers dynamics of perturbations in the $P(X)$-theory: $\omega^2 =c^2_s k^2$, where $c_s$ is the sound speed in the $P(X)$-theory. We refer to the associated modes and the dispersion relation as hydrodynamical ones. Note that the $P(X)$-theory describes a perfect fluid for the timelike $\partial_{\mu} \varphi$. Hence, the name ''hydrodynamical''.

The behavior of the other d.o.f. (which we dub as the non-hydrodynamical one) depends on the 
sound speed $c_s$ of the corresponding $P(X)$-theory. These perturbations are stable in the subluminal case, $c_s<1$, see Sec.~\ref{subsec:Subluminality}. 
Moreover, the spectrum of the non-hydrodynamical modes is separated by a mass gap from that of the hydrodynamical modes. The mass gap depends on the form of the potential of the complex field, and can be 
large enough, so that an observer interested only in low momentum dynamics does not see the high frequency non-hydrodynamical modes. In other words, from the point of view of a low energy effective field theory dynamics of the complex scalar field is indistinguishable from that of the $P(X)$-theory. 

Interestingly, there is a class of potentials leading to superluminal hydrodynamical modes,  $c_s>1$ (Sec.~\ref{subsec:Superluminality}). Nevertheless, information in perturbations of a canonical complex scalar field cannot propagate faster than light, since the front velocity is unity. Indeed, as discussed above the dispersion relation is standard for high momenta, $\omega^{2}={k}^{2}$. That is, the appearance of superluminality is an artefact of working in the low momentum limit. Contrary to the subluminal case, the non-hydrodynamical d.o.f. is plagued by a tachyonic instability.  
This instability compromises the relation with the $P(X)$-theory. 

In Sec.~\ref{subsec:GradientUnstable}, we consider canonical complex scalar field models corresponding to the $P(X)$-theories with $c^2_s<0$. The latter are plagued by a catastrophic gradient instability. Models of the complex field provide with a low energy cutoff ameliorating this instability and making such $P(X)$-theories more physically interesting. 

In Sec.~\ref{sec:Inflation}, we also discuss a mechanism of generating a particular configuration of the complex field mimicking the $P(X)$-theory. This is possible in the specific model, where the phase of the field $\Psi$ is coupled to the inflaton. The resulting background profile of the field $\Psi$ is characterized by the nearly constant amplitude---one of the 
necessary conditions for recovering the $P(X)$-theory. We also find that adiabatic and 
isocurvature modes of the field $\Psi$ correspond to hydrodynamical and non-hydrodynamical modes, respectively. 
Hence, suppressing isocurvature perturbations, one automatically suppresses non-hydrodynamical perturbations of the 
field $\Psi$. 

\section{From $P(X)$-theory to a complex scalar field\label{sec:From-fluid-to-U(1)}}
To see the link between the $P(X)$-theory and the model of the complex
scalar field, let us consider the following representation of the
$P(X)$-theory Lagrangian\footnote{In this Section, our discussion closely follows Ref.~\cite{Babichev:2017lrx}.} 
\begin{equation}
\mathscr{L}=\chi{}^{2} X-V(\chi) \;,\label{actionpx}
\end{equation}
 where 
\begin{equation}
\nonumber
X\equiv\frac{1}{2}(\partial\varphi)^{2}\;,
\end{equation}
 is the standard kinetic term of the field $\varphi$. The latter is assumed to be dimensionless, while the auxiliary field $\chi$ is dimensionful. The equation of motion for the field $\chi$ is given by
\begin{equation}
\frac{V_{\chi}}{\chi}=2X \label{constraint}\; .
\end{equation}
Hereafter the subscript $'\chi'$ denotes the derivative with respect to the field $\chi$.
 Expressing the field $\chi$ as the function of $X$ and plugging
back into the action, one obtains some generic Lagrangian which is a function of $X$. We
end up with the $P(X)$-theory.

The stress-energy tensor of the $P(X)$-theory reads 
\begin{equation}
\label{stresspx}
T_{\mu \nu} =\chi^{2} \partial_{\mu}\varphi\partial_{\nu}\varphi -g_{\mu\nu}
\left( \frac{\chi}{2}V_{\chi}-V \right) \; ,
\end{equation}
where $\chi$ is the subject to the constraint~\eqref{constraint}. For the time-like $\partial_{\mu} \varphi$ the $P(X)$-theory describes
a perfect fluid with energy density and pressure given by 
\begin{equation}
\varepsilon=\frac{\chi}{2}V_{\chi}+V \; , \label{energylambda}
\end{equation}
and 
\begin{equation}
p=\frac{\chi}{2}V_{\chi}-V \; . \label{pressurelambda}
\end{equation}
Comparing these expressions with the standard results for the $P(X)$-theory one obtains\footnote{Hence, in this way one can only describe $P(X)$-theories, which satisfy the Null Energy Condition, cf. Ref.~\cite{deRham:2017aoj}.} 
\begin{equation}
\label{pxchirelation}
\chi^2 =P_{X} \; .
\end{equation}
Small perturbations in the $P(X)$-perfect fluid propagate with the sound speed given by the standard expression\footnote{In the cosmological context this formula was obtained
in Ref.~\cite{Garriga:1999vw}.}
\begin{equation}
c_{s}^{2}=\frac{\partial p}{\partial\varepsilon}=\left(1+\frac{2XP_{XX}}{P_{X}}\right)^{-1}  \label{sspx}\; ;
\end{equation}
the subscript $'X'$ denotes the derivative with respect to $X$.
 In terms of the field $\chi$ this expression can be rewritten as
\begin{equation}
c_{s}^{2}=\frac{M_{2}^{2}}{M_{1}^{2}}\;,\label{soundpx}
\end{equation}
 where we introduced the shorthand notations: 
\begin{equation}
M_{1}^{2}\equiv V_{\chi \chi}+3\frac{V_{\chi}}{\chi} \quad \text{and}\quad M_{2}^{2}\equiv V_{\chi \chi}-\frac{V_{\chi}}{\chi }\;.\label{masses}
\end{equation}
$M_{1}$ and $M_{2}$ are two parameters (generically time-dependent)
of the mass dimension. 

Now, let us promote the field $\chi$ to the dynamical d.o.f.
by adding the kinetic term~\cite{Babichev:2017lrx}: 
\begin{equation}
\nonumber
\frac{1}{2}(\partial\chi)^{2}\;,
\end{equation}
 to the Lagrangian \eqref{actionpx}. With this extra term, the Lagrangian takes the form: 
\begin{equation}
\mathscr{L}=\frac{1}{2}\left(\left(\partial\chi\right)^{2}+\chi^{2}\left(\partial\varphi\right)^2\right)-V\left(\chi\right) \; . \label{eq:Lagrangian_polar}
\end{equation}
The latter can be rewritten as the Lagrangian of the canonical complex scalar field $\Psi=\chi e^{i\varphi}$,
\begin{equation}
\label{actioncomplex}
\mathscr{L}=\frac{1}{2}|\partial\Psi|^{2}-V(|\Psi|) \; .
\end{equation}
We arrive at the model of the canonical
globally $U(1)$-charged complex scalar field with some
self-interacting potential. The corresponding stress-energy tensor is given by 
\begin{equation}
T_{\mu\nu}=\partial_{\mu}\chi\partial_{\nu}\chi+\chi^{2} \partial_{\mu}\varphi\partial_{\nu}\varphi-g_{\mu\nu}\mathscr{L}\,. \label{eq:EMT_polar2}
\end{equation}
The resulting equations of motion for the amplitude $\chi$ and the phase $\varphi$ are 
\begin{equation}
\label{eqampl}
\square \chi -\chi (\partial \varphi )^2+V_{\chi}=0\; , 
\end{equation}
and 
\begin{equation}
\label{eqphase}
\nabla_{\mu} \left(\chi^2 \nabla^{\mu} \varphi \right) =0\; ,
\end{equation}
respectively. 

Dynamics of the complex field models is richer compared to that of the $P(X)$-theories.
The reason is the extra d.o.f. encoded in the field $\chi$. For the complex field model to reproduce the $P(X)$-theory, dynamics of $\chi$
should remain frozen until the times, when caustic singularities are
supposed to be formed. At the times of the (would be) caustics formation,
the extra d.o.f. comes into play and smoothens caustics. 

Two qualifications are in order here. For any time-like $\partial \varphi$, one has $c^2_s=1$ if and only if $P_{XX}=0$. In terms of the field $\chi$, 
the second derivative $P_{XX}$ reads 
\begin{equation}
\nonumber 
P_{XX}=\frac{4\chi^3}{\chi V_{\chi \chi}-V_{\chi}} \; .
\end{equation}
Hence, for any regular potential, $P_{XX}=0$ implies $\chi=0$ where the description in terms of two fields $\chi$ and $\varphi$ as in Eq.~\eqref{actionpx} breaks down. 
Indeed, the field $\varphi$ is ill-defined in the limit $\chi \rightarrow 0$. We conclude that the equivalence of the $P(X)$-theories and the models~\eqref{actionpx} is no longer valid in the limit $c^2_s \rightarrow 1$. 

Second, the procedure of completing
by means of the complex scalar field formally can be applied to the ghost
condensate $P(X)\propto(X-\Lambda^2)^{2}$, where $\Lambda$ is some dimensionful parameter, see Ref. \cite{ArkaniHamed:2003uy}. 
However, for the ghost condensate one
is interested in dynamics of the field $\varphi$ around $P_X=0$, or in other terms around $\chi=0$, see Eq.~\eqref{pxchirelation}.
But the phase $\varphi$ of the complex scalar is ill-defined in that case. Therefore, the completion by the complex field is not applicable to the ghost condensate at the point of condensation\footnote{However, away from the condensation point such a completion can be constructed \cite{Bilic:2008pe}.}.
\section{Homogeneous cosmology}\label{sec:HomoCosmo}

As we have discussed in the previous Section, the $P(X)$-theory is reproduced from the model~\eqref{actioncomplex} 
provided that the amplitude $\chi$ gets frozen. Up to the Hubble drag such a configuration of the complex field can be easily achieved in the homogeneous cosmology. Consider the equation of motion~\eqref{eqphase} for the phase $\varphi$, which reduces to 
\begin{equation}
\label{phasebackground}
\ddot{\varphi}+\left(3H+2\frac{\dot{\chi}}{\chi}\right)\dot{\varphi}=0 \; .
\end{equation}
This equation corresponds to the conservation of
the U(1)-Noether charge
\begin{equation}
\label{nc} 
Q=\chi^{2} \dot{\varphi}\; ,
\end{equation}
in the comoving
volume, i.e., 
\begin{equation}
\label{ncshift}
Q=\frac{C}{a^{3}}\;,
\end{equation}
 where $C$ is some constant and $a$ is the scale factor. The equation of motion for the amplitude~\eqref{eqampl} reduces to   
\begin{equation}
\label{eqamplcosmology}
\ddot{\chi}+3H\dot{\chi}-\chi \dot{\varphi}^{2}+V_{\chi}=0\; .
\end{equation}
The $P(X)$-theory is recovered at the background level, provided that the first two terms here are negligible. The latter condition is exactly what 
we mean by freezing out the amplitude $\chi$. Then, the solution of Eq.~\eqref{eqamplcosmology} reads\footnote{With no loss of generality, we choose the velocity of the phase to
be positive.}
\begin{equation}
\dot{\varphi}=\sqrt{\frac{V_{\chi}}{\chi}}\;.\label{phaseback}
\end{equation}
Note that the existence of this solution requires that 
\begin{equation}
V_{\chi}>0\;.\label{backgroundstability}
\end{equation}
Under certain conditions, Eq.~\eqref{phaseback} is indeed consistent with our initial assumption of the amplitude $\chi$ being frozen out, and thus serves as an approximate 
solution of the complex field models.

Let us show this explicitly. Taking the time derivative of Eq.~\eqref{phaseback} and using it again one obtains 
\begin{equation}
\frac{\ddot{\varphi}}{\dot{\varphi}}=\frac{1}{2}\left(\frac{M_{2}}{\dot{\varphi}}\right)^{2}\,\frac{\dot{\chi}}{\chi}\; .
\label{Key_Relation}
\end{equation}
Using this relation one can either exclude  $\varphi$ from the equation of motion for the phase \eqref{phasebackground} and get 
\begin{equation}
\label{amplitudeslow}
\dot{\chi}=-\frac{3}{2} H\chi  (1-c^2_s) \; ,
\end{equation}
or exclude $\dot{\chi}$ instead, and obtain 
\begin{equation}
\ddot{\varphi}+3c_{s}^{2}H\dot{\varphi}=0\; ,\label{eq:phase_EoM}
\end{equation}
where $c^2_s$ is given by Eq.~\eqref{soundpx}. 
The last equation of motion is identical to that of the corresponding $P(X)$-theory. From~\eqref{amplitudeslow} it follows that the amplitude $\chi$ is decreasing for $c^2_s<1$ and grows for $c^2_s>1$, while \eqref{eq:phase_EoM} implies that $\dot{\varphi}$ redshifts for any real non-zero $c_s$. Both equations can be easily integrated for constant $c_s$ 
 \begin{equation}
\label{phasevelred}
\chi\propto a^{-3\left(1-c_{s}^{2}\right)/2} \; , \quad\text{and}\quad \dot{\varphi}\propto a^{-3c_{s}^{2}} \; .
\end{equation}
We do not know yet the meaning of the quantity $c_s$ in the model of the complex scalar. 
As we will see later, it plays the role of the phase velocity for the low frequency perturbations of the complex field. Here it is important that 
$c^2_s$ is naturally not much larger than unity. Then from Eq. \eqref{amplitudeslow} it follows that the first two terms in Eq.~\eqref{eqamplcosmology} are indeed parametrically small, provided that $H$ and $c_s$ are slowly changing quantities and the following condition is satisfied
\begin{equation}
\frac{V_{\chi}}{\chi}\gg H^{2}\;.\label{condition}
\end{equation}
By making use of Eq.~\eqref{phaseback}, this can be rewritten as the condition on the phase time derivative 
\begin{equation}
\label{condition2}
\dot{\varphi} \gg H \; .
\end{equation}
Note that for the background~\eqref{phaseback}, the mass parameters $M_1$ and $M_2$ defined by Eq.~\eqref{masses} 
can be expressed as the functions of $\dot{\varphi}$ and the quantity $c_s$: 
\begin{equation}
\label{massvarphi} 
M^2_1=\frac{4\dot{\varphi}^2}{1-c^2_s} \; , \quad\text{and}\quad M^2_2 =\frac{4c^2_s\dot{\varphi}^2}{1-c^2_s} \; .
\end{equation}
Hence, for $c^2_s \sim 1$ and $1- c^2_s = {\cal{O}}(1)$, the time derivative $\dot{\varphi}$ and the masses $M_1,~M_2$ are of the same order of magnitude, 
$\dot{\varphi} \sim M_1 \sim M_2$. 

Notably, there is clear physical motivation underlying the solution~\eqref{phaseback}. Using Eq.~\eqref{nc} we can exclude $\dot\varphi$ from Eq.~\eqref{eqamplcosmology}: 
\begin{equation}
\label{single_dof}
\ddot{\chi}+3H\dot{\chi}-\frac{Q^{2}}{\chi^{3}}+V_{\chi}=0\;.
\end{equation}
Hence, the amplitude $\chi$ evolves in the effective potential 
\begin{equation}
\label{V_eff}
V_{eff}=\frac{Q^{2}}{2\chi^{2}}+V \; .
\end{equation}
Its minimum is located at 
\begin{equation}
\label{min_eff}
{\chi}=\left(\frac{Q^{2}}{V_{\chi}({\chi})}\right)^{1/3}\;,
\end{equation}
which exactly corresponds to \eqref{phaseback}, see also Ref.~\cite{Tolley:2009fg}. We conclude that dynamics of the complex field with the amplitude $\chi$ set at the minimum 
of its effective potential can be described in terms of the $P(X)$-theories.

From Eq.~\eqref{phasevelred} it follows that the velocity of the phase $\dot{\varphi}$ is constant modulo the Hubble drag. Provided
that the condition~\eqref{condition} is satisfied, one can show that
the stress-energy tensor associated with the complex field $\Psi$ given by Eq.~\eqref{eq:EMT_polar2} is
that of the $P(X)$-theory given by Eq.~\eqref{stresspx}. Indeed, from Eqs.~\eqref{amplitudeslow} and~\eqref{condition2} one gets $\dot{\chi}^{2}\ll\chi^{2} \dot{\varphi}^{2}$. Hence the equality of the stress-energy tensors. Recall that the homogeneous Universe is assumed here. Linear perturbations will be considered in the next Section.

\section{Dispersion relations\label{sec:Dispersion-relations-Cosmo}}

When studying linear perturbations of the complex scalar, we discard metric fluctuations. We assume that the cosmological modes characterized by the conformal momentum ${\bf k}$ are in the sub-horizon regime, i.e., $k/a \gg H$. Neglecting terms of the order $a^2H^2/k^2$ and 
switching to the conformal time defined from $ad\eta=dt$, one writes linearized equations for the amplitude and the phase
\begin{equation}
\label{eqlambda} 
\delta''_{\chi}+[3c^2_s-1]{\cal H} \delta'_{\chi}+\left[k^2+M^2_2a^2 \right]\delta_{\chi}=2\varphi' \, \delta \varphi' \; , 
\end{equation}
and
\begin{equation}
\label{eqvarphi}
\delta \varphi''+[3c^2_s-1]{\cal H} \delta \varphi'+k^2 \delta \varphi =-2\varphi' \, \delta '_{\chi} \; ,
\end{equation}
where $\delta_{\chi} \equiv \delta \chi /\chi$ and $\delta \varphi$ are the Fourier modes of the relative amplitude and the phase perturbations with the conformal momentum ${\bf k}$.  
The prime denotes the derivative with respect to the conformal time, and ${\cal H}=a'/a$. Recall that the mass parameters $M_{1}$ and $M_{2}$ are defined
by Eq.~\eqref{masses}. Writing Eqs.~\eqref{eqlambda} and Eq.~\eqref{eqvarphi}, we made use of the background equations for the phase~\eqref{phaseback} and the amplitude~\eqref{amplitudeslow}. 
Again we keep $c^2_s$ given by Eq.~\eqref{soundpx} as a shorthand notation for the ratio $M^2_2/M^2_1$ not assuming any physical meaning behind it at the moment. 
Below we solve Eqs.~\eqref{eqlambda} and~\eqref{eqvarphi} in the WKB approximation and confirm the results by a numerical analysis, see Fig.~\ref{fig:modes}. 

Two comments are in order here. First, for a quartic potential, which models radiation, Eqs.~\eqref{eqlambda} and~\eqref{eqvarphi} can be solved exactly for an arbitrary expansion history $H(t)$, see appendix \ref{sec:phi4}. Second, there is a physically 
interesting analogy for our system. We observe that Eqs.~\eqref{eqlambda} and~\eqref{eqvarphi} 
describe the motion of a damped charged oscillator on a two dimensional plane immersed in a strong orthogonal magnetic field. Motivated by this analogy, in appendix~\ref{sec:Average} we solve equations of motion for perturbations by averaging over rapid oscillations.

Following WKB method, we decompose the phase and the amplitude perturbations as follows: 
\begin{equation}
\label{wkb}
\delta_{\chi} =\alpha  \cdot e^{ i \int \omega d\eta  -\int \gamma_1 {\cal H} d\eta }\; , \quad\text{and}\quad \delta \varphi =\beta \cdot e^{i \int \omega d\eta -\int \gamma_2 {\cal H} d \eta }  \; .
\end{equation}
Here $\alpha$ and $\beta$ are the constant amplitudes, $\omega$ is the frequency, and $\gamma_1$ and $\gamma_2$ are two dimensionless functions taking order one values, which parametrize the decay/growth of perturbations 
in the expanding Universe. We assume that $\omega$, $\gamma_1$, and $\gamma_2$ are changing slowly on the time scale $\sim \omega^{-1}$. 
Furthermore, their time dependence is only due to the Hubble drag, so that $ \omega'/\omega \sim \gamma'_1/\gamma_1 \sim \gamma'_2/\gamma_2 \sim {\cal H}$. 
In particular, these two conditions imply $\omega \gg {\cal H}$. 

Substituting Eqs.~\eqref{wkb} into Eqs.~\eqref{eqlambda} and~\eqref{eqvarphi} and omitting the terms of the order ${\cal H}^2/k^2$ and ${\cal H}^2/\omega^2$, one obtains
\begin{equation}
\label{wkbb}
\left[k^2-\omega^2+M^2_2a^2+i\omega'+i \left[3c^2_s-1-2\gamma_1 \right]\omega {\cal H}\right] \cdot \alpha -2\varphi' \left[i\omega - \gamma_2{\cal H} \right] 
e^{\int [\gamma_1-\gamma_2] {\cal H} d \eta } \cdot \beta =0\; ,
\end{equation}
and
\begin{equation}
\label{wkba}
2\varphi' \left[i\omega - \gamma_1 {\cal H}\right]  e^{\int [\gamma_2-\gamma_1] {\cal H} d \eta } \cdot \alpha+\left[k^2-\omega^2  +i\omega'+i [3c^2_s-1-2\gamma_2] \omega {\cal H} \right] \cdot \beta=0 \; .
\end{equation}
We result with the system of homogeneous equations with respect to the constants $\alpha$ and $\beta$. 
It has the non-trivial solution provided that its determinant equals to zero.

Both real and imaginary parts of the determinant should be set to zero. The former condition results into the biquadratic equation defining the frequency $\omega$, 
\begin{equation}
\label{biquadratic}
\omega^{4}-\omega^{2}\left(2k^2+M_{1}^{2}a^2\right)+M_{2}^{2}a^2k^{2}+k^{4}=0\;.
\end{equation}
The solution to Eq.~\eqref{biquadratic} reads  
\begin{equation}
\omega_{\pm}^{2}=k^{2}+\frac{1}{2}M_{1}^{2}a^2\pm \frac{a}{2}\sqrt{M_1^4a^2+4k^2(M_1^2-M_2^2)} \; , \label{exact} 
\end{equation}
cf.~\cite{Tsumagari:2009na,Boyle:2001du,Achucarro:2010jv,Achucarro:2010da}. Hereafter the uppescripts $''+''$ and $''-''$ correspond to
the choice of the positive and negative sign in Eq.~\eqref{exact},
respectively. In the high momentum limit, $k \rightarrow\infty$, one immediately
obtains the standard dispersion relation 
\begin{equation}
\nonumber
{\omega}_{\pm}^{2}=k^{2}\; ,
\end{equation}
 as one could expect. Further analysis of Eq.~\eqref{exact} will be performed in the next Subsections.

The equality to zero of the imaginary part of the determinant gives 
\begin{equation}
\label{cond1}
\left[3c^2_s-1+\frac{d \ln \omega }{d \ln a}-2\gamma_2 \right] M^2_2 a^2 -4\varphi'^2(\gamma_1+\gamma_2)+2\left[3c^2_s-1+\frac{d \ln \omega}{d \ln a}-\gamma_1-\gamma_2\right] (k^2 -\omega^2)=0 \; .  
\end{equation}
The extra condition determining the functions $\gamma_1$ and $\gamma_2$ comes from the requirement that the amplitudes $\alpha$ and $\beta$ are constant. 
We extract the ratio $\alpha/\beta$ from Eq.~\eqref{wkbb} and neglect the terms suppressed by the Hubble rate ${\cal H}$, 
\begin{equation}
\label{rat}
\frac{\alpha}{\beta}=\frac{2i\omega \varphi'}{k^2-\omega^2+M^2_2a^2} e^{\int (\gamma_1-\gamma_2) {\cal H} d \eta }  \; .
\end{equation}
Taking the derivative of the left and right hand sides with respect to $\ln a$, we obtain
\begin{equation}
\label{cond2}
\gamma_1-\gamma_2=-\frac{d}{d\ln a} \ln \frac{\omega \varphi'}{k^2-\omega^2+M^2_2a^2} \; .
\end{equation}
In what follows, we will show that the dynamical properties of the complex field models encoded in Eqs.~\eqref{exact},~\eqref{cond1}, and~\eqref{cond2} 
indeed match those of the $P(X)$-theory. 

\subsection{Correspondence to $P(X)$-theories: subluminal case\label{subsec:Subluminality}}

When studying the correspondence to the $P(X)$-theories, we first ignore the effects due to the cosmic expansion and focus on the dispersion relation~\eqref{exact}. 
The stability considerations require that both frequencies $\omega_{+}$ and $\omega_{-}$ following from Eq.~\eqref{exact} are real. We will see that in the low momentum limit defined as 
\begin{equation}
\label{hierarchy} 
k \ll {\rm{min}\{|M_1|,|M_2|\}} \cdot a \; ,
\end{equation}
this condition is fulfilled, when matching the complex scalar model to the subluminal
$P(X)$-theory. On the other hand, in the case of the superluminal
$P(X)$-theories there is always a tachyonic instability present. Therefore,
it makes sense to consider these two cases separately.

First, let us consider the case 
\begin{equation}
M_{1}^{2}>0\; , \label{ineqone}
\end{equation}
where $M_1$ is defined by Eq.~\eqref{masses}. Choosing the branch with the negative sign in Eq.~\eqref{exact} and assuming the low momentum limit, one gets
\begin{equation}
{\omega}_{-}^{2}=\frac{M_{2}^{2}}{M_{1}^{2}} \cdot {k}^{2}+{\cal O}\left(\frac{k^{4}}{a^2M^2_1} \right)\; ,\label{negativesub}
\end{equation}
cf. Refs.~\cite{Tolley:2009fg,Achucarro:2010jv,Achucarro:2010da}. The same with the positive sign in Eq.~\eqref{exact} reads 
\begin{equation}
{\omega}_{+}^{2}=M_{1}^{2}a^2+\frac{2M_{1}^{2}-M_{2}^{2}}{M_{1}^{2}}{k}^{2}+{\cal O} \left(\frac{{k}^{4}}{a^2M^2_1} \right)\;.\label{positivesub}
\end{equation}
Note that the quartic correction $\sim k^4$ stems from the terms $\sim k^4$ and $\omega^2 k^2$ in Eq.~\eqref{biquadratic}. 

The physical frequencies $\omega^{ph}$ are obtained from the frequencies $\omega$ defined with respect to the conformal time by a trivial rescaling
\begin{equation}
\nonumber 
\omega^{ph}_{-}\approx \frac{M_2}{M_1} \cdot \frac{k}{a} \; , \qquad \omega^{ph}_{+} \approx M_1 \; .
\end{equation} 
The absence of gradient instabilities for the branch~\eqref{negativesub} imposes the condition 
\begin{equation}
M_{2}^{2}>0\;,\label{nogradientsub}
\end{equation}
which restricts the choice of the potentials $V(\chi)$.

Now, comparing Eq.~\eqref{negativesub} with Eq.~\eqref{soundpx}, we see
that the branch with the negative sign exactly reproduces the dispersion relation in the $P(X)$-theory. Furthermore, from Eqs.~\eqref{masses} and~\eqref{backgroundstability} it is clear that $M^2_1>M^2_2$, which for positive 
$M^2_1$ and $M^2_2$ implies that  
\begin{equation}
\nonumber
c_{s}^{2}=\frac{M_{2}^{2}}{M_{1}^{2}}<1\;.
\end{equation}
Hence, we deal with subluminal perturbations. It is important that $\omega_{+}^{2}$
is positive. Consequently, the complex field model reproducing the subluminal $P(X)$-theory is stable. We see that in the low momentum limit the frequencies $\omega_{-}$ and $\omega_{+}$ are separated by the large mass gap set by the mass 
parameter $M_1$. Therefore, for the device with the resolution threshold well below both $M_1$ and $M_2$ the complex field perturbations are indistinguishable from those 
in the $P(X)$-theory. In particular, the $k^4$-term in \eqref{negativesub} is negligible.  

However, for the parametrically small speed of sound $c_{s}\ll1$ there is another interesting range of momenta  
\begin{equation}
M_{2}\ll k/a\ll M_{1}\;, 
\end{equation}
or equivalently  
\begin{equation}
\label{condition_dispersion}
c_{s}\varphi'\ll k\ll\varphi'\;,
\end{equation}
see Eq.~\eqref{massvarphi}. With the $k^4$-term written explicitly, Eq.~\eqref{negativesub} takes the form   
\begin{equation}
\label{diff_disper}
\omega_{-}^{2}=c_{s}^{2}k^{2}+\frac{\left(1-c_{s}^{2}\right)^{3}}{4\varphi'^{2}}k^{4}+...
\end{equation}
Here we again used \eqref{massvarphi}. In this range of momenta the $k^4$-term in the dispersion relation \eqref{diff_disper} is dominant over the standard hydrodynamical one\footnote{Note that the $k^6$-term is still negligible.}. Hence the phase velocity deviates from the sound speed of the $P(X)$-theory and substantially depends on the frequency. We postpone a detailed discussion of the evolution of the amplitudes of perturbations in this regime to Appendix \ref{sec:Average}. 

\subsection{Correspondence to $P(X)$-theories: superluminal case\label{subsec:Superluminality} }
Now, consider the case 
\begin{equation}
M_{1}^{2}<0\;.\label{ineqsuper}
\end{equation}
Then Eqs.~\eqref{negativesub} and~\eqref{positivesub} are still true modulo the replacement
\begin{equation}
\nonumber
\omega_{+}^{2}\leftrightarrow\omega_{-}^{2}\;.
\end{equation}
 The absence of gradient instabilities imposes the condition 
\begin{equation}
\nonumber
M_{2}^{2}<0\;,
\end{equation}
 which is automatically satisfied, once the inequalities~\eqref{backgroundstability}
and~\eqref{ineqsuper} are obeyed. Looking at Eq.~\eqref{soundpx},
we see that the dispersion relation for $\omega_{+}^{2}$ is that
of the $P(X)$-theory. Using Eqs.~\eqref{backgroundstability} and~\eqref{ineqsuper}
one can show that the corresponding sound speed squared is larger than unity,  
\begin{equation}
\nonumber
c_{s}^{2}=\frac{M_{2}^{2}}{M_{1}^{2}}>1\;.
\end{equation}
Hence, we are recovering a superluminal $P(X)$-theory. However, contrary to the superluminality considered in \cite{Babichev:2007dw}, here the front of any wave still propagates with the speed of light. Note that the superluminality requires concave potentials. Indeed, from the definition of $M^2_1$ in Eq.~\eqref{masses} and the inequality~\eqref{ineqsuper} it 
follows that $V_{\chi \chi} <-3 V_{\chi}/\chi<0$. 

There is a problem, however: the superluminal case is plagued by the tachyon instability. Indeed, consider the non-hydrodynamical branch of Eq.~\eqref{exact}, the one with the negative sign. It gives in the low momentum limit: 
\begin{equation}
\nonumber 
{\omega}^2_{-} =M^2_1a^2+{\cal O} ({k}^2)<0 \; .
\end{equation} 
 Note that this problem is related but complementary to those pointed out in \cite{Adams:2006sv}\footnote{For a recent discussion see Ref.~\cite{Chandrasekaran:2018qmx}.}. The tachyon instability
is developed at the time scales $\tau_{tach}\simeq|M_{1}|^{-1}$. For small $\tau_{tach}$, the tachyon instability is
very fast, and the system quickly decays to the stable state. Instead,
if $\tau_{tach}$ is comparable (but smaller) than the age of the
Universe $\sim H_{0}^{-1}$, superluminality can be in principle observed
at very long wavelengths of the order of the present horizon. For
even larger $\tau_{tach}$, superluminality does not pop out --- all
cosmologically relevant modes have the standard dispersion relation $\omega^{2}=k^{2}$. 

Note that transitions from superluminal configurations to subluminal ones and vice versa are impossible in our basic approximation \eqref{phaseback}. Indeed, as we mentioned at the end of Sec.~\ref{sec:From-fluid-to-U(1)}, they cannot occur through $c_s=1$. The only remaining possibility would be a jump from $c^2_s \rightarrow +\infty$ to $c^2_s \rightarrow -\infty$, taking place as $M^2_1$ crosses zero, while $M^2_2<0$, see Eq.~\eqref{masses}. However, this divergence in the sound speed invalidates our approximation, see Eq.~\eqref{amplitudeslow}. 

\subsection{Completing gradient unstable $P(X)$-theories\label{subsec:GradientUnstable}}
So far we have avoided discussing $P(X)$-theories with the negative sound speed 
squared, $c^2_s<0$. The reason is the presence of gradient instabilities, which invalidate the scenarios of interest, 
unless there is a cutoff on the momenta of unstable modes. Luckily such a cutoff is provided by the models of the complex field\footnote{See Refs.~\cite{Garcia-Saenz:2018vqf,Garcia-Saenz:2018ifx} for the inflationary setups, which lead to the infrared gradient instabilities.}. 

We assume that the non-hydrodynamical modes are stable, so that $M^2_1>0$. Then, the low frequency modes describing perturbations of the complex scalar are given 
by Eq.~\eqref{diff_disper}.
Here $c^2_s$ is again given by Eq.~\eqref{soundpx}. We assume that $M^2_2<0$, so that $c^2_s<0$. Note that the potential can be still convex, $V_{\chi \chi}>0$. 

As $c^2_s <0$, the first term on the r.h.s. of Eq.~\eqref{diff_disper} 
indicates the instability. Note that this case is also plagued by ghosts, as it follows from Eq. (C.6) of Ref. \cite{Babichev:2007dw}. For relatively high momenta, however, both instabilities are regularized because the second term on the r.h.s. is positive. 
The cutoff on the unstable modes is given by 
\begin{equation}
\nonumber 
k_s \simeq |c_s| M_1a\simeq |c_s|\varphi'\; .
\end{equation} 
Hence, the maximal rate of instability is estimated as $\Gamma \simeq |\omega_s|/a \simeq |c^2_s| \dot \varphi$. We keep $c^2_s$ as a small parameter, i.e., $|c^2_s| \ll 1$, so that $\Gamma \ll M_1$. 
Furthermore, one can choose the range of parameters $c_s$ and $M_1$ in order to make the rate $\Gamma$ smaller than the present Hubble rate.
\subsection{Effects of cosmic expansion}
\label{subsec:expansion}
Now let us include the effects of the cosmic expansion into the analysis. These are encoded in the functions $\gamma_1$ and $\gamma_2$ defined by Eq.~\eqref{wkb}. 
One determines $\gamma_1$ and $\gamma_2$ from Eqs.~\eqref{cond1} and~\eqref{cond2}. To simplify these relations, recall that the spectrum of the complex field perturbations has two branches. 
Only the low frequency one is of interest for us, because it gives the dispersion relation matching that of the $P(X)$-theories. Therefore, in Eqs.~\eqref{cond1} and~\eqref{cond2} we 
take the limit $\omega \ll \varphi', M_1,~M_2$, so that Eq.~\eqref{cond2} simplifies to 
\begin{equation}
\nonumber 
\gamma_1-\gamma_2=-\frac{d }{d\ln a} \ln \frac{\omega \varphi'}{M^2_2a^2} \; .
\end{equation}  
Substituting $\omega=c_s k$ and making use of Eqs.~\eqref{phasevelred},~\eqref{massvarphi}, one obtains 
\begin{equation}
\label{rel}
\gamma_1-\gamma_2=1-3c^2_s +\frac{1+c^2_s}{1-c^2_s}\frac{d\ln c_s}{d \ln a}\; .
\end{equation} 
Eqs.~\eqref{cond1} and~\eqref{rel} are sufficient in order to define the functions $\gamma_1$ and $\gamma_2$. The latter reads  
\begin{equation}
\label{decay}
\gamma_2=\frac{3c^2_s-1}{2} -\frac12\frac{d\ln c_s}{d\ln a}\; .
\end{equation} 
Plugging this into Eq.~\eqref{wkb}, one obtains the time dependence of the phase perturbations
\begin{equation}
\label{lowfrphaseexp}
\delta \varphi \propto \sqrt{c_s} e^{-\frac12\int(3c^2_s-1)d\ln a} e^{i\int c_sk d\eta} \; .
\end{equation}
For the constant sound speed $c_s$, perturbations $\delta \varphi$ grow provided that $c^2_s<1/3$ and redshift if $c^2_s >1/3$. 
For $c^2_s=1/3$ phase perturbations oscillate with the constant amplitude---this is the consequence of the scale symmetry emerging in that case. 

\begin{figure}
\centering
\includegraphics[width=\textwidth]{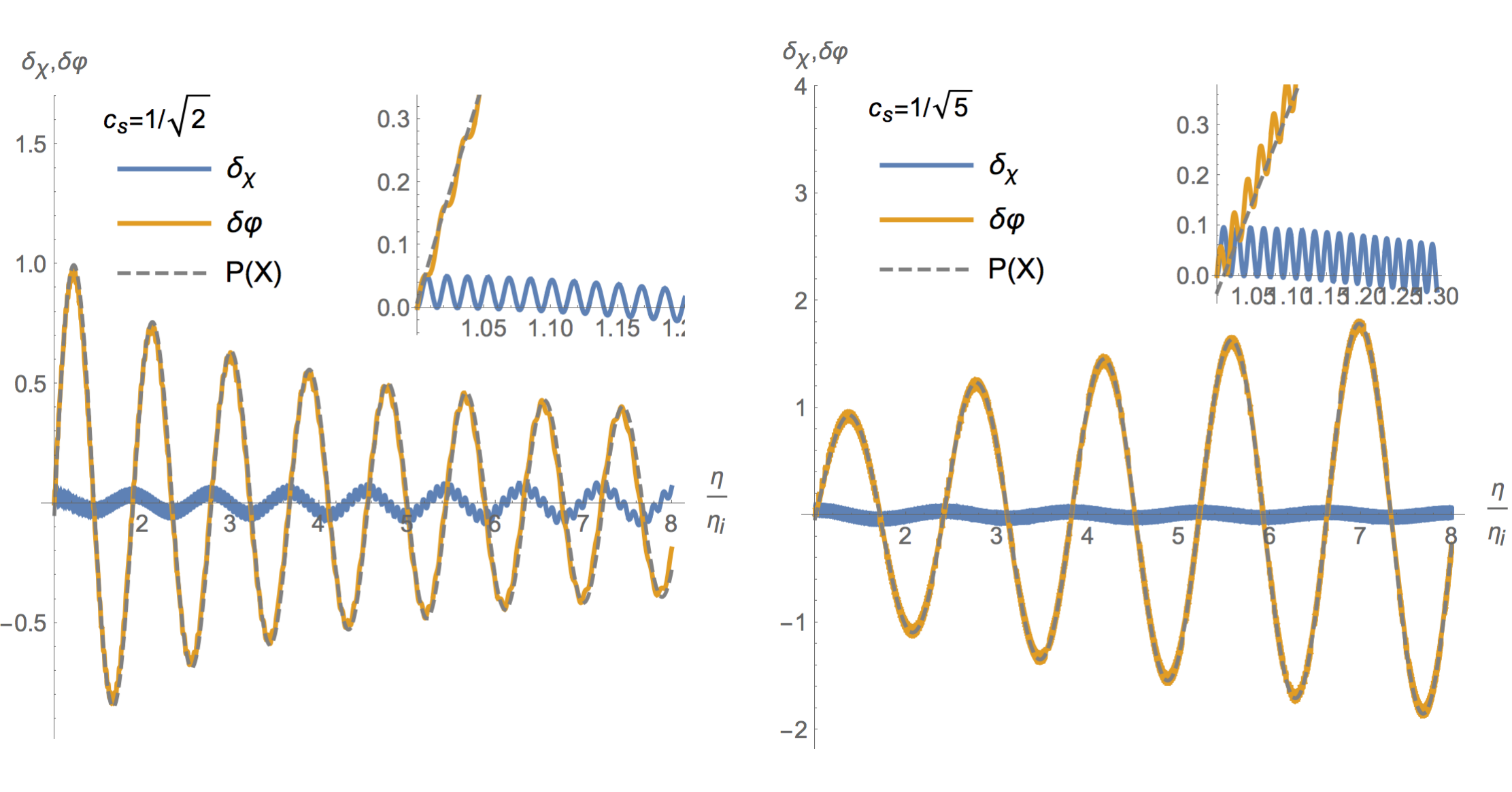}
\caption{The conformal time evolution of perturbation modes in the complex field models and in the $P(X)$-theories is shown for two cases: $c_s^2=1/2$  and $c_s^2=1/5$. 
Matter dominated Universe with the Hubble rate ${\cal H}=2/\eta$ is assumed. Solid blue and orange lines depict the amplitude and phase perturbations, $\delta_{\chi}$ 
and $\delta \varphi$, in the complex field models; the dashed line describes perturbations of the $P(X)$-field $\varphi$. Each mode of the complex scalar contains low frequency and high frequency contributions, as it can be seen from the subplots.
There is a perfect agreement in the behavior of the low frequency modes of the phase perturbations and the $P(X)$-field perturbations. 
We have chosen the following initial values to fulfil the conditions formulated in Sec.~\ref{subsec:Subluminality} and~\ref{subsec:expansion}: $k\eta_i = 10$ (modes are in the sub-horizon regime) and  $\varphi'_i/k=15$ (guarantees the existence of low frequency modes with the hydrodynamical dispersion relation, see Eq.~\eqref{hierarchy}), where the index $i$ denotes the initial values taken at $\eta=\eta_i$. Note that perturbations $\delta\varphi$ decay for $c_s^2>1/3$ (left panel) and grow for $c_s^2<1/3$ (right panel), 
in an agreement with our calculation in Sec.~\ref{subsec:expansion}. 
}
\label{fig:modes}
\end{figure}

Finally, let us determine the time dependence of the amplitude perturbations $\delta_{\chi}$. From Eqs.~\eqref{rel} and~\eqref{decay}, we obtain
\begin{equation}
\nonumber 
\gamma_1 =-\frac{3c^2_s-1}{2} +\frac{3c^2_s+1}{2(1-c^2_s)} \frac{d\ln c_s}{d\ln a} \; .
\end{equation}
Substituting this into Eq.~\eqref{wkb} and integrating over the conformal time $\eta$, one gets
\begin{equation}
\label{lowframplexp}
\delta_{\chi}  \propto \frac{|1-c_s^2|}{\sqrt{c_s}} e^{\frac12\int(3c^2_s-1)d\ln a} e^{i\int c_sk d\eta} \; .
\end{equation}
Note that for the low frequency branch of the spectrum, $\omega \ll M_1,~M_2$, amplitude perturbations $\delta_{\chi}$ are suppressed compared to the phase ones $\delta{\varphi}$. This follows from Eq.~\eqref{rat}, 
\begin{equation}
\label{hie}
\frac{\delta_{\chi}}{\delta \varphi} \sim  \frac{\alpha}{\beta}   \sim \frac{i \omega}{\varphi'} \; , 
\end{equation}
where, as usual, we assumed no hierarchy between $\varphi'$ and $M_2$. We also observe that the amplitude perturbations $\delta_\chi$ are shifted by the phase $\pi/2$ relative to the phase ones $\delta\varphi$.
Both these features can be seen in Fig.~\ref{fig:modes}.

In Appendix \ref{sec:Average} we give an alternative derivation of the above results.  There we also cover the special case of a parametrically small speed of sound. 

Let us contrast the expression~\eqref{lowfrphaseexp} to the behavior of the $P(X)$-field $\varphi$ perturbations. We use the standard representation of the $P(X)$-theory, i.e., not involving the auxiliary field $\chi$. 
The equation of motion for the field $\varphi$ is then given by
\begin{equation}
\nonumber 
\nabla_{\mu} \left(P_{X} \nabla^{\mu} \varphi \right) =0 \; .
\end{equation}
Linearizing the latter, we obtain 
\begin{equation}
\nonumber 
\delta \varphi''+\left[(3c^2_s-1){\cal H} -2\frac{c'_s}{c_s} \right]\delta \varphi' +c^2_s k^2 \delta \varphi=0 \; .
\end{equation}
Here we made use of the expression~\eqref{sspx}. We again substitute phase perturbations in the form $\delta \varphi=\beta e^{i \int \omega d \eta-\int \gamma_2 {\cal H} d\eta}$. Neglecting terms of the order ${\cal H}^2$, we obtain 
\begin{equation}
\nonumber 
-\omega^2+c^2_s k^2+i\omega \left[(3c^2_s-1) {\cal H} -2\gamma_2 {\cal H} +\frac{\omega'}{\omega} -\frac{2c'_s}{c_s} \right]=0 \; .
\end{equation}
Both real and imaginary parts of that equation should equal to zero. The former condition gives the dispersion relation $\omega^2 =c^2_s k^2$, as is expected. The latter condition defines
 the function $\gamma_s$ as in Eq.~\eqref{decay}. 
We conclude that the perturbations in $P(X)$-theory 
indeed have the same time dependence as the phase perturbations of the complex scalar field in the low frequency regime.

We checked our analytical expressions~\eqref{lowfrphaseexp} and~\eqref{lowframplexp} by numerically solving Eqs.~\eqref{wkbb} and~\eqref{wkba} for different values of the sound speed $c_s$, which we kept constant, and compared them with the evolution of perturbations of the $P(X)$-field $\varphi$. The results are shown in Fig.~\ref{fig:modes}.


\section{$P(X)$-theory from inflation}\label{sec:Inflation}
We observed in the previous Sections that the $P(X)$-theory can be
completed by means of the complex scalar for the particular configuration of the latter. Here we discuss the mechanism of classically
producing the complex scalar from inflation with initial conditions, which automatically yield such configurations. Our discussion in this Section parallels to that of Ref.~\cite{Babichev:2018afx},
and we stress on the essential points below. 

Consider the following coupling of the phase of the complex scalar
to the inflaton 
\begin{equation}
\nonumber
S_{int}=\beta \int d^{4}x\sqrt{-g}\cdot\varphi\cdot T_{infl}\;.
\end{equation}
 Here $T_{infl}$ is the trace of the inflaton energy-momentum tensor and $\beta$
is some dimensionless constant. This interaction explicitly violates $U(1)$-symmetry and hence leads to the generation of the Noether charge density estimated by 
\begin{equation}
\nonumber
Q\equiv\chi^{2} \dot{\varphi} \simeq\frac{\beta U}{H}\;,
\end{equation}
 where $U$ is the inflaton potential, and $H$ is the Hubble
rate during inflation. In the presence of the non-zero Noether charge
density $Q$, the equation of motion for the homogeneous amplitude
$\chi$ is given by \eqref{single_dof}. The amplitude $\chi$ evolves in the effective potential \eqref{V_eff}. For sufficiently steep potentials satisfying $M^2_1,~M^2_2 \gtrsim H^2$, the field $\chi$ relaxes to its minimum \eqref{min_eff} within a few Hubble times.
We see that the amplitude ${\chi}$ of the complex scalar generated
from inflation is nearly constant. It is exactly constant in the de
Sitter space-time approximation and small variations are measured by slow roll parameters. 

Now let us consider perturbations of the fields $\chi$ and $\varphi$ during inflation. As usual, we split 
perturbations into adiabatic (which are due to the inflaton) and isocurvature ones (which the field $\Psi$ has on its own). These have been studied in Ref.~\cite{Babichev:2018afx} for the 
case of the free massive complex scalar field. The generalization to the case of the self-interacting potentials is straightforward. Below we list the main results. The super-horizon 
adiabatic perturbations $\delta \chi_{ad}$ and $\delta \varphi_{ad}$ (the subscript 'ad' stands for 'adiabatic') obey
\begin{equation}
\nonumber
\frac{\delta\chi_{ad}}{\chi'}=\frac{\delta\varphi_{ad}}{\varphi'}=\frac{\delta\phi}{\phi'}\; ,
\end{equation}
where $\phi$ is the inflaton field. Again we switch to the conformal time, when studying perturbations. We see that the perturbations $\delta\chi_{ad}$ and $\delta\varphi_{ad}$
remain nearly constant behind the horizon during inflation. 

When discussing isocurvature perturbations, one can set inflaton fluctuations
as well as metric fluctuations to zero. Then, the equations for $\delta_{\chi, iso} \equiv \delta \chi_{iso}/\chi$
and $\delta\varphi_{iso}$ (the subscript 'iso' stands for 'isocurvature') are given by
\begin{equation}
\nonumber
\delta''_{\chi, iso}+2{\cal H}\delta'_{\chi, iso}-\partial_{i}\partial_{i}\delta_{\chi, iso}+M^2_2a^2 \delta_{\chi, iso}-2\varphi' \delta \varphi_{iso}=0\;,\label{systemone}
\end{equation}
 and 
\begin{equation}
\delta \varphi''+2{\cal H} \delta \varphi'+2\delta'_{\chi, iso} \varphi'=0\;.
\label{systemoneone}
\end{equation}
We assume the constant background for the amplitude $\chi$, i.e., $\chi=const$. Being interested in the super-horizon regime, we neglect spatial derivatives of the fields. Then Eq.~\eqref{systemoneone} simplifies to
\begin{equation}
\nonumber
\left[a^2 \delta (\chi^2 \varphi') \right]'=0\; .
\end{equation}
 Consequently, one gets
\begin{equation}
\nonumber
\delta(\chi^{2}\varphi'_{iso})=\frac{C}{a^{2}}\; ,
\end{equation}
where $C$ is the integration constant. The r.h.s. here redshifts fast during inflation, and we obtain 
\begin{equation}
\frac{\delta \varphi'_{iso}}{\varphi'}=-2\delta_{\chi, iso}\;.\label{corr}
\end{equation}
We use the latter to express the perturbation $\delta \varphi'$. Plugging it back into Eq.~\eqref{systemone}, we get
\begin{equation}
\nonumber
\delta''_{\chi, iso}+2{\cal H}\delta'_{\chi, iso}+ M_{1}^{2}a^2\delta_{\chi, iso}=0\; .
\end{equation}
 Recall that for the positive $M_{1}^{2}$, the complex scalar field perturbations reproduce those of the $P(X)$-theory 
modulo the non-hydrodynamical perturbations with the frequency $M_1$. Now we see that the latter can be identified as isocurvature perturbations (they have the same frequency $M_1$). Furthermore, provided that $M^2_1a^2 \gtrsim {\cal H}^2$,
 the isocurvature modes decay fast in the super-horizon regime. Consequently, non-hydrodynamical modes are not excited in the 
spectrum of complex field perturbations.

To summarize, for $M^2_1, M^2_2 \gtrsim H^2$, initial conditions set by inflation 
correspond to the configuration of the complex field with the properties of the 
subluminal $P(X)$-theory. Instead, for negative $M_{1}^{2}$, the isocurvature perturbations are plagued by a tachyon instability. 
Recall that negative $M^2_1$ correspond to the superluminal $P(X)$-theory. Hence, there is no natural way to obtain the 
superluminal $P(X)$-theory from inflation. 

\section{Discussions}

In the present work, we discussed the possibility of completing 
$P(X)$-theories by means of the self-interacting canonical complex scalar field. 
Generically, a completion is necessary because $P(X)$-theories 
develop caustics and have obscure quantum properties. On the flipside, 
the canonical scalar is manifestly free of caustics; furthermore, there is a known prescription for its quantization, at least for 
renormalizable potentials. 

We have shown that 
the correspondence between subluminal $P(X)$-theories and the complex field models indeed holds in cosmology 
assuming the proper background configuration of the complex field. 
A ``proper'' background configuration is such that both the amplitude of the complex scalar field $\chi$ and the phase time derivative $\dot\varphi$ are constant modulo the Hubble drag. 
This happens when $\dot\varphi$ is large in comparison to the Hubble rate, Eq.~(\ref{condition2}). 

We have shown in Sec~\ref{subsec:Subluminality} that the low energy spectrum of the complex scalar perturbations coincides with that of the subluminal $P(X)$-theory. The correspondence between the complex scalar models and the $P(X)$-theories breaks down at $k/a\sim c_s \dot\varphi$, where the $k^4$-correction to the dispersion relation becomes relevant, Eq.~\eqref{diff_disper}. For subluminal theories, the other, non-hydrodynamical, branch of the spectrum of perturbations contains high energy modes separated by a mass gap $\sim\dot\varphi$, see Eq.~\eqref{positivesub}.
Thus from the effective field theory point of view dynamics of the complex scalar at low energies is fully described by the $P(X)$-theory.
In addition we have shown in Sec.~\ref{sec:Inflation} that the high energy modes can be suppressed in the inflationary framework. 

It is in principle possible to have a theory with the low energy superluminal modes, $c_s>1$, see Sec.~\ref{subsec:Superluminality}.
In this case, however, the non-hydrodynamical branch of the spectrum of the complex scalar 
contains a tachyonic instability.
This case still can be of interest, if the time of instability is of the order of the age of the Universe. 

We also discussed a subclass of complex scalar field models with gradient instabilities at low momenta, Sec.~\ref{subsec:GradientUnstable}.
For $P(X)$-theories a case with $c_s^2<0$ leads to catastrophic instabilities. 
The complex scalar field provides a regularization of the gradient instabilities at high momenta. Namely, in the complete picture 
physical momenta  $k /a \gtrsim |c_s| \dot\varphi$ are stable and have the non-relativistic dispersion relation $\omega \propto k^2$. 

In Sec.~\ref{subsec:expansion} we included the effects of the cosmic expansion on the evolution of the perturbations.
We demonstrated that the low frequency perturbations of the phase of the complex scalar field evolve in a full agreement with perturbations of the $P(X)$-theory.

We finalize with some concluding remarks and prospects for the future. The results obtained in this paper may have implications for modified gravity models, especially in the light of the recent detection of the gravitational signal GW170817 
and its counterpart GRB170817A~\cite{TheLIGOScientific:2017qsa,Monitor:2017mdv}. This observation tightly constrained the speed of gravitational waves to be very close to the speed of light. 
Based on this, one is inclined to rule out a large class of interesting modified gravity models, where gravitational waves do not propagate with the speed of light.
However, this conclusion might be erroneous for the following reason. 
From the effective field theory point of view, a modified gravity model has a cutoff $\Lambda$ which may be lower than the energy scale of gravitational waves observed by LIGO (e.g., in~\cite{deRham:2018red} it was argued that the energy scales observed at LIGO are very close to the cutoff). 
At the same time, a completion of this model at high energies may have the speed of gravity equal to unity,---in a comfortable agreement with the data.
In this paper we provide an exactly solvable toy model with such a behavior. Indeed, for momenta lower than $\dot\varphi$, the speed of scalar perturbations is $c_s\neq 1$; while at high energies the canonical dispersion relation $\omega^2=k^2$ is recovered.

While we focused on the cosmological backgrounds in the present work, the correspondence 
to the $P(X)$-theories can be extended to inhomogeneous 
and anisotropic backgrounds. We consider this generalization in a forthcoming paper where 
we also develop a hydrodynamical description of the complex scalar field models.

\acknowledgments\label{Ackno} 

A.V. thanks Lasha Berezhiani, Pavel Kovtun, Ignacy Sawicki, and Dam Thanh
Son for useful discussions and criticism. It is a pleasure to thank Grant Remmen,
S\'ebastien Renaux-Petel, and Andrew Tolley for a useful correspondence.  
E.B. acknowledges support from PRC CNRS/RFBR (2018--2020) n\textsuperscript{o}1985 ``Gravit\'e modifi\'ee et trous noirs: signatures 
exp\'erimentales et mod\`eles consistants'' and 
from the research program ``Programme national de cosmologie et galaxies'' of the CNRS/INSU, France. 
The work of S.R. and A.V. was supported by the funds from the European Regional Development
Fund and the Czech Ministry of Education, Youth and Sports (M\v{S}MT):
Project CoGraDS - CZ.02.1.01/0.0/0.0/15\_003/0000437. A.V. also acknowledges
support from the J. E. Purkyn\v{e} Fellowship of the Czech Academy
of Sciences. 

\appendix
\section{Perturbations for quartic potential}
\label{sec:phi4}
In the main part of the text we solved equations of motion for perturbations \eqref{eqlambda} and \eqref{eqvarphi} using the WKB method. Here we discuss a particular case when they can be solved exactly.  Let us consider an instance of the potential $V\propto\chi^{4}$. The correspondence \eqref{energylambda}, \eqref{pressurelambda}, \eqref{soundpx} and \eqref{masses} implies that this system models radiation with $c_{s}^{2}=1/3$. In this physically interesting case 
the friction terms disappear from the equations of motion for perturbations (\ref{eqlambda}) and (\ref{eqvarphi}). 
Furthermore the equation of motion for the background phase (\ref{eq:phase_EoM}) 
\[
\varphi''+\left(3c_{s}^{2}-1\right)\mathcal{H}\varphi'=0\; ,
\label{eq:conformal_phi}
\]
implies that for radiation $\varphi'=const$, and consequently 
\begin{equation}
\nonumber
a^{2}M_{2}^{2}=2\varphi'^{2}\text=const \; ;
\end{equation}
(cf.~Eq.~\eqref{massvarphi}). Hence, equations of
motion (\ref{eqlambda}) and (\ref{eqvarphi}) for perturbations
$\delta_{\chi}$ and $\delta \varphi$ build a system of ordinary linear differential equations with \emph{time-independent} coefficients 
\begin{align}
 &  \delta_{\chi}''+\left(k^{2}+2\varphi'^{2}\right)\delta_{\chi}=2\varphi'\,\delta\varphi' \; ,\label{eq:System_Perturb_radiation}\\
 &\delta\varphi''+k^{2}\delta\varphi=-2\varphi'\,\delta_{\chi}'\; . \nonumber 
\end{align}
As a consequence, dynamics of perturbations is identical for all cosmological backgrounds, $H(t)$. 
We have already mentioned in the main text that this system describes the motion of an anisotropic and charged oscillator on a 2d plane $\left(\delta_{\chi},\varphi\right)$ immersed in a strong orthogonal magnetic field with the cyclotron frequency $2\varphi' $, see appendix \ref{sec:Average}. 

To solve Eqs.~\eqref{eqlambda} and~\eqref{eqvarphi}, we apply the ansatz 
\begin{equation}
\delta_{\chi}=\alpha e^{i\omega\eta}\;  \mbox{and}~\delta\varphi=\beta e^{i\omega\eta}\; ,\label{eq:substitute_rad_perturb}
\end{equation}
and obtain
\begin{align*}
 & \left(k^{2}-\omega^{2}+2\varphi'^{2}\right)\alpha-2i\omega\varphi'\beta=0\; ,\\
 & 2i\omega\varphi'\alpha+\left(k^{2}-\omega^{2}\right)\beta=0\; .
\end{align*}
This system has a non-trivial solution provided that the determinant of
the corresponding matrix is vanishing, so that 
\begin{equation}
\nonumber
\omega^{4}-\omega^{2}\left(2k^{2}+6\varphi'^{2}\right)+k^{2}\left(k^{2}+2\varphi'^{2}\right)=0\; .
\end{equation}
This biquadratic equation has two solutions 
\begin{equation}
\nonumber
\omega_{\pm}^{2}=k^{2}+3\varphi'^{2}\pm\sqrt{9\varphi'^{4}+4k^{2}\varphi'^{2}}\; ,
\end{equation}
cf.~Eq.~\eqref{exact} and Ref.~\cite{Alford:2012vn}. The general solution reads
\begin{equation}
\nonumber
\delta_{\chi}=\alpha_{+}e^{i\omega_{+}\eta}+c_{+}e^{-i\omega_{+}\eta}+\alpha_{-}e^{i\omega_{-}\eta}+c_{-}e^{-i\omega_{-}\eta}\; ,
\end{equation}
and 
\begin{equation}
\nonumber
\delta\varphi=\frac{2i\omega_{+}\varphi'}{\omega_{+}^{2}-k^{2}}\left[\alpha_{+}e^{i\omega_{+}\eta}-c_{+}e^{-i\omega_{+}\eta}\right]+\frac{2i\omega_{-}\varphi'}{\omega_{-}^{2}-k^{2}}\left[\alpha_{-}e^{i\omega_{-}\eta}-c_{-}e^{-i\omega_{-}\eta}\right]\; ,
\end{equation}
where $\left(\alpha_{+},\alpha_{-},c_{+},c_{-}\right)$ are four independent complex constants. 

For $k^{2}\ll\varphi'^{2}$ the effective magnetic field is crucial for dynamics leading to the strong violation of the canonical dispersion relation $\omega^2 =k^2$. 
The dispersion relations for the hydrodynamical and non-hydrodynamical modes read
\begin{equation}
\nonumber
\omega_{-}^{2}=\frac{1}{3}k^{2}+\frac{2 k^4}{27\varphi'^{2}}+...\; ,
\end{equation}
and
\begin{equation}
\nonumber
\omega_{+}^{2}\simeq6\varphi'^{2}+\frac{5}{3}k^{2}+...\; ,
\end{equation}
respectively. 

Finally, we note that four independent amplitudes in $\delta_{\chi}$ governing modes with $\omega_{-}$ and $\omega_{+}$ can be of the same order. This is not the case of perturbations $\delta\varphi$, since 
\begin{equation}
\nonumber
\frac{2\omega_{+}\varphi'}{\omega_{+}^{2}-k^{2}}\simeq\sqrt{\frac{2}{3}}\;, \quad\text{and}\quad\frac{2\omega_{-}\varphi'}{\omega_{-}^{2}-k^{2}}\simeq-\sqrt{3}\left(\frac{\varphi'}{k}\right)\; .
\end{equation}
Therefore for $k\ll\varphi'$ and $\alpha_{+} \sim \alpha_{-} \sim c_{-} \sim c_{+}$ we have that the hydrodynamical mode, with $\omega_{-}$, is dominant in $\delta\varphi$ and enhanced by the factor $\varphi'/k$. 
\section{Averaging over oscillations in magnetic field}
\label{sec:Average}
In this appendix we discuss a physically interesting analogy for our equations~\eqref{eqlambda} and~\eqref{eqvarphi}. 
Namely, the same system describes the motion of a 
damped\footnote{Expansion of the Universe only leads to actual damping for $c_{s}^{2}>1/3$, otherwise it works as an antidamping.} anisotropic charged oscillator on a two dimensional  plane immersed in a strong orthogonal magnetic field $B$\footnote{Alternatively, one can consider a damped anisotropic oscillator moving in a rotating frame, so that the gyroscopic force on the r.h.s. of Eqs.~\eqref{eqlambda} and \eqref{eqvarphi} 
corresponds to the Coriolis force. In that case the angular velocity of the rotating frame maps as $\Omega\rightarrow\varphi'$, cf. problem 3,~\cite[p.~129]{Landafshitz_Mechanics}. However, to obtain equations of motion \eqref{eqlambda} and \eqref{eqvarphi} one has to neglect the centrifugal force what makes this analogy less consistent.}. One just replaces $2\varphi'$ on the r.h.s. of Eqs.~\eqref{eqlambda} and~\eqref{eqvarphi} by the cyclotron frequency, $\omega_{c}=eB/mc$. The oscillator has a unit mass $m$ and spring constants $k^{2}$ in one direction and $(k^{2}+M^2_2a^2)$
in another direction, cf.~\cite[p.~59]{Landafshitz_Teorpol}\footnote{There is a typo in this English edition of Landau and Lifshitz Vol .2: a wrong sign on the r.h.s. of the equation of motion for $y$.}. Using this analogy, we find another way of solving Eqs.~\eqref{eqlambda} and~\eqref{eqvarphi}. The method is based on averaging of the high frequency oscillations.

Indeed, we can write equations of motion~\eqref{eqlambda} and~\eqref{eqvarphi} for perturbations,
$\mathbf{q}=\left(\delta_{\chi},\delta\varphi\right)$, in the form
of the Lagrange equations with the Lorentz and dissipative forces
\begin{equation}
\frac{d}{d\eta}\frac{\partial L}{\partial\mathbf{v}}-\frac{\partial L}{\partial\mathbf{q}}=\mathbf{F}_{L}+\mathbf{F}_{H}\; .\label{eq:Lagrange_Eq_F}
\end{equation}
The Lagrange function here describes two oscillators 
\begin{equation}
L=\frac{1}{2}\left(\delta\varphi'^{2}-k^{2}\delta\varphi^{2}\right)+\frac{1}{2}\left(\delta_{\chi}'^{2}-\left(k^{2}+a^{2}M_{2}^{2}\right)\delta_{\chi}^{2}\right)\; ; \label{eq:Lagrange_Function}
\end{equation}
the dissipative force $\mathbf{F}_{H}$ caused by the Hubble drag is given by 
\begin{equation}
\mathbf{F}_{H}=-\left(3c_{s}^{2}-1\right)\mathcal{H}\mathbf{v}\; , \label{eq:friction_Force}
\end{equation}
while the Lorentz force $\mathbf{F}_{L}$ is 
\begin{equation}
\mathbf{F}_{L}=e\,\mathbf{v}\times\mathbf{B}\; , \label{eq:Lorentz_Force}
\end{equation}
where the magnetic field is orthogonal to $\mathbf{q}$ and has the
absolute value $eB=2\varphi'$. The presence of this gyroscopic force does not allow to find normal modes.  

Now let us discuss the averaging method of solving Eqs.~\eqref{eqlambda} and~\eqref{eqvarphi}. As for the first step, we neglect the Hubble drag and make use of the ansatz~\eqref{eq:substitute_rad_perturb}. We obtain 
\begin{align}
 & \left(k^{2}+a^{2}M_{2}^{2}-\omega^{2}\right)\alpha-2i\omega\varphi'\beta=0\; ,\label{eq:linear_for_magnetic}\\
 & \left(k^{2}-\omega^{2}\right)\beta+2i\omega\varphi'\alpha=0\; . \nonumber 
\end{align}
The requirement that the determinant of this system vanishes gives
the dispersion relation~\eqref{exact} from the main text. In the low momentum limit, the hydrodynamical modes have the dispersion relation $\omega_{-}\simeq c_{s}k$, so that 
\begin{equation}
\beta=\alpha\cdot\frac{2i\omega_{-}\varphi'}{\omega_{-}^{2}-k^{2}}\simeq-\alpha\cdot\frac{2ic_{s}}{1-c_{s}^{2}}\cdot\frac{\varphi'}{k}\; , \label{eq:alpha_betta_ratio}
\end{equation}
and we conclude that $\beta\gg\alpha$, cf.~\eqref{rat}. The Lorentz force is gyroscopic and does not
change the energy, so that 
\begin{equation}
E=\mathbf{v}\frac{\partial L}{\partial\mathbf{v}}-L=\frac{1}{2}\left(\delta\varphi'^{2}+k^{2}\delta\varphi^{2}\right)+\frac{1}{2}\left(\delta_{\chi}'^{2}+\left(k^{2}+a^{2}M_{2}^{2}\right)\delta_{\chi}^{2}\right)\; . \label{eq:Energy}
\end{equation}
Now let us switch on the Hubble expansion. Due to the dissipative force
and the time dependence of the spring constant the energy is not conserved,
rather it changes in accordance with 
\begin{equation}
\frac{dE}{d\eta}=\mathbf{v}\mathbf{F}_{H}-\frac{\partial L}{\partial\eta}\; , \label{eq:Energy_Nonconservation}
\end{equation}
where 
\begin{equation}
\frac{\partial L}{\partial\eta}=-\frac{1}{2}\delta_{\chi}^{2}\left(a^{2}M_{2}^{2}\right)'\; ,\label{eq:partial_L}
\end{equation}
and 
\begin{equation}
\mathbf{v}\mathbf{F}_{H}=-\left(3c_{s}^{2}-1\right)\mathcal{H}\left(\delta\varphi'^{2}+\delta_{\chi}'^{2}\right)\; . \label{eq:dissipative_power}
\end{equation}
We plug in the solution for the hydrodynamic mode and average over many oscillations assuming that the
Hubble drag is very weak and the change of the Lagrangian is very slow.
We also promote the constants $\alpha$ and $\beta$ to slowly
changing variables. 

On average we have 
\begin{equation}
\overline{\delta_{\chi}^{2}}=\frac{1}{2}\overline{\alpha^{2}} \; ,\quad\overline{\delta_{\chi}'^{2}}=\frac{\omega_{-}^{2}}{2}\overline{\alpha^{2}}\; , \quad\text{and}\quad\overline{\delta\varphi^{2}}=\frac{1}{2}\overline{\beta^{2}}\; , \quad\overline{\delta\varphi'^{2}}=\frac{\omega_{-}^{2}}{2}\overline{\beta^{2}}\; , \label{eq:elementary_average}
\end{equation}
where $\omega_{-}\simeq c_{s}k$. Now we plug these expressions in the average energy, use Eq.~(\ref{eq:alpha_betta_ratio})
to eliminate $\alpha$ and Eq.~\eqref{eq:conformal_phi} to express $\varphi''$. Provided that the following inequalities hold,
\begin{equation}
\mathcal{H}\ll c_{s}k\ll c_{s}^{2}\varphi'\; , \label{eq:condition_nontrivial}
\end{equation}
(cf.~Eqs.~\eqref{hierarchy} and~\eqref{condition_dispersion}), we obtain for the averaged quantities in the leading order 
\begin{equation}
\overline{E}=\frac{1}{2}k^{2}\beta^{2}\; ,\label{eq:average_energy}
\end{equation}
\begin{equation}
\overline{\frac{\partial L}{\partial\eta}}=-\frac{1}{2}k^{2}\beta^{2}\left[\frac{c_{s}'}{c_{s}}-\left(3c_{s}^{2}-1\right)\left(1-c_{s}^{2}\right)\mathcal{H}\right]\; , \label{eq:average_change_Lagrangian}
\end{equation}
and 
\begin{equation}
\overline{\mathbf{v}\mathbf{F}_{H}}=-\frac{\left(3c_{s}^{2}-1\right)}{2}\mathcal{H}c_{s}^{2}k^{2}\beta^{2}\; .\label{eq:average_dissipative_power}
\end{equation}
The time derivative of the energy is 
\begin{equation}
\frac{d\overline{E}}{d\eta}=k^{2}\beta\beta' \label{eq:derivative_Ebar} \; .
\end{equation}
Inserting these averaged quantities into the energy evolution
equation \eqref{eq:Energy_Nonconservation} one obtains 
\begin{equation}
\frac{\beta'}{\beta}=\frac{1}{2}\frac{c_{s}'}{c_{s}}-\frac{\left(3c_{s}^{2}-1\right)}{2}\mathcal{H}\; .\label{eq:result_betta}
\end{equation}
Making a trivial integration, one obtains $\beta\propto\exp\left(-\int d\eta\gamma_{2}\mathcal{H}\right)$, cf.~ Eq.~\eqref{wkb}, where $\gamma_2$ is given by Eq.~\eqref{decay}, consistently 
with our calculation in the main text.

Suppose that the sound speed $c_{s}\ll1$. Then, the dispersion relations
\eqref{exact} and \eqref{diff_disper} give for the gapless mode
\begin{equation}
\nonumber
\omega_{-}^{2}\simeq k^{2}\left(c_{s}^{2}+\frac{k^{2}}{4\varphi'^{2}}+\frac{k^{4}}{8\varphi'^{4}}+...\right)\; .
\end{equation}
Hence, for the range of the wavenumbers
\begin{equation}
c_{s}\varphi'\ll k\ll\varphi'\; , \label{eq:range_nonrelativistic}
\end{equation}
the dispersion relation is non-relativistic 
\begin{equation}
\omega_{-}\simeq\frac{k^{2}}{2\varphi'}\; .\label{eq:non_relativist_dispers}
\end{equation}
The waves we consider should be inside the Hubble scale, $\omega_{-}\gg\mathcal{H}$,
so that 
\begin{equation}
k\gg\sqrt{\varphi'\mathcal{H}}\; . \label{eq:nonrelativistic_inside}
\end{equation}
The non-relativistic dispersion relation (\ref{eq:non_relativist_dispers})
yields for the amplitudes \eqref{eq:substitute_rad_perturb}:
\begin{equation}
\beta=\alpha\cdot\frac{2i\omega_{-}\varphi'}{\omega_{-}^{2}-k^{2}}\simeq-i\alpha\; , \label{eq:amplitudes_nonrelativist}
\end{equation}
so that both fluctuations $\delta\varphi$ and $\delta_{\chi}$ have the same order of magnitude, contrary to the case $k\ll c_{s}\varphi'$, see Eq.~\eqref{eq:alpha_betta_ratio}. 

Using the relation between the amplitudes (\ref{eq:amplitudes_nonrelativist})
and averaged perturbations~\eqref{eq:elementary_average} with the non-relativistic
dispersion relation (\ref{eq:non_relativist_dispers}) one obtains
the same expression \eqref{eq:average_energy} for the average energy. At the same time, the averaged power of the Hubble drag~\eqref{eq:dissipative_power}
is given by 
\begin{equation}
\overline{\mathbf{v}\mathbf{F}_{H}}=\frac{1}{4}\frac{k^{4}\mathcal{H}}{\varphi'^{2}}\beta^{2}\; . \label{eq:Hubble_power_nonrelativ}
\end{equation}
The averaged time derivative of the Lagrangian is given by
\begin{equation}
\overline{\frac{\partial L}{\partial\eta}}=-2\beta^{2}c_{s}^{2}\varphi'^{2}\left(\mathcal{H}+\frac{c'_{s}}{c_{s}}\right)\; . \label{eq:Lagrangian_evolve_nonrel}
\end{equation}
Now we substitute expressions \eqref{eq:average_energy},~\eqref{eq:derivative_Ebar},~\eqref{eq:Hubble_power_nonrelativ}, and~\eqref{eq:Lagrangian_evolve_nonrel} into Eq.~\eqref{eq:Energy_Nonconservation}
and obtain 
\begin{equation}
\frac{\beta'}{\beta}=\frac{\mathcal{H}}{4}\left(\frac{k^{2}}{\varphi'^{2}}+\frac{8c_{s}^{2}\varphi'^{2}}{k^{2}}\right)+\frac{2c_{s}^{2}\varphi'^{2}}{k^{2}}\cdot\frac{c'_{s}}{c_{s}}\; .\label{eq:Amplitude_evolution_Nonrelat}
\end{equation}
For the wavenumbers 
\begin{equation}
k\gg\sqrt{c_{s}}\varphi'\; , \label{eq:dust_scales}
\end{equation}
the power of the Hubble drag is dominant and the evolution of the
amplitudes is independent on $c_{s}$ in the leading order
\begin{equation}
\frac{\beta'}{\beta}=\frac{\mathcal{H}}{4}\left(\frac{k}{\varphi'}\right)^{2}\; , \label{eq:evolution_for_dust_scales}
\end{equation}
which reads in terms of $\gamma_2$ as follows 
\begin{equation}
\nonumber
\gamma_2=-\frac{1}{4}\left(\frac{k}{\varphi'}\right)^2\; .
\end{equation}
Hence these scales evolve as in the system with the vanishing $c_{s}$. 

However, for the wavenumbers in the range 
\begin{equation}
c_{s}\varphi'\ll k\ll\sqrt{c_{s}}\varphi'\; , \label{eq:intermediate_scales}
\end{equation}
the time derivative of the Lagrangian is stronger than the power of the Hubble
drag, so that the evolution of the amplitude is described by 
\begin{equation}
\frac{\beta'}{\beta}=\frac{2c_{s}^{2}\varphi'^{2}}{k^{2}}\left(\mathcal{H}+\frac{c'_{s}}{c_{s}}\right)\; ,\label{eq:modes_evolv_intermediate} \; .
\end{equation}
In terms of $\gamma_2$ this reads as follows 
\begin{equation}
\nonumber
\gamma_2=-\frac{2c_{s}^{2}\varphi'^{2}}{k^{2}}\left(1+\frac{d \ln c_{s}}{d \ln a}\right)\; .
\end{equation}
To sum up, for a small sound speed there are four different regimes
inside the Hubble horizon: 
\begin{enumerate}
\item Long wavelength perturbations, $\mathcal{H}\ll c_{s} k \ll c_{s}^{2}\varphi'$ have the same hydrodynamical dispersion relation as 
the perturbations in the $P(X)$-theories. Perturbations $\delta_{\chi}$ are parametrically suppressed compared to phase perturbations $\delta \varphi$. This regime is only possible provided that
\[
c_{s}^{2}\gg\frac{\mathcal{H}}{\varphi'}\; .\label{eq:cond_for_hydro}
\]
\item Intermediate, shorter, wavelength perturbations with $c_{s}\varphi'\ll k\ll\sqrt{c_{s}}\varphi'$
have the non-relativistic dispersion relation (\ref{eq:non_relativist_dispers}) and thus are different from those in the corresponding $P\left(X\right)$-theories. 
Amplitudes $\alpha$ and $\beta$ of perturbations $\delta_{\chi}$ and $\delta \varphi$, respectively, have the 
same order of magnitude, as it follows from Eq.~(\ref{eq:amplitudes_nonrelativist}). Their
evolution is still affected by the non-vanishing sound speed (\ref{eq:modes_evolv_intermediate}). These modes are inside the horizon, provided that Eq.~\eqref{eq:nonrelativistic_inside} holds, i.e., in the range $\sqrt{\varphi'\mathcal{H}}\ll k\ll\sqrt{c_{s}}\varphi'$. As it follows, this regime is only possible for 
\begin{equation}
\nonumber
c_{s}\gg\frac{\mathcal{H}}{\varphi'}\; ,
\end{equation}
which is weaker than the condition~\eqref{eq:cond_for_hydro}.
\item For even shorter wavelengths with $\sqrt{c_{s}}\varphi'\ll k\ll\varphi'$ 
the dispersion relation is again non-relativistic, Eq.~(\ref{eq:non_relativist_dispers}), and thus 
is different from that in the $P\left(X\right)$-theories. The amplitudes $\alpha$ and $\beta$ have the same 
order of magnitude, Eq.~(\ref{eq:amplitudes_nonrelativist}), but now their evolution is
independent on the sound speed (\ref{eq:evolution_for_dust_scales}). 
\item For short, ultraviolet, wavelengths with $k\gg\varphi'$ both modes of the complex scalar
field $\Psi$ propagate with the speed of light. 
\end{enumerate}
The same range of scales appears also in the case of infrared gradient instabilities,
$c_{s}^{2}<0$. The formulas obtained above are applicable to this case up to an
obvious replacement $c_{s}\rightarrow\left|c_{s}\right|$, where necessary. 

\bibliographystyle{utphys}
\addcontentsline{toc}{section}{\refname}\bibliography{U1_Fluid}

\end{document}